\documentclass[aps,reprint,longbibliography,superscriptaddress]{revtex4-1}
\usepackage{amsmath}
\usepackage{amssymb}
\usepackage{bm}
\usepackage{epsfig}
\usepackage{graphicx}
\usepackage[dvipsnames]{xcolor}
\usepackage{color}
\usepackage[unicode=true,colorlinks=true,citecolor=blue,urlcolor=blue]{hyperref}
\usepackage[T2A]{fontenc}
\usepackage[cp1251]{inputenc}
\usepackage[english]{babel}
\usepackage{braket}

\usepackage{ulem}
\usepackage{multirow}

\graphicspath{{img/}}
\usepackage{tikz}
\usetikzlibrary{arrows.meta}
\usetikzlibrary{decorations.pathmorphing,decorations.markings}

\newcommand{\arctanh}{\mathop{\rm arctanh}}

\newcommand{\e}{\mathrm{e}}

\renewcommand{\i}{{\rm i}}
\renewcommand{\d}{\mathrm d}
\renewcommand{\emph}{\textit}

\usepackage{amsthm}

\renewcommand{\vec}[1]{{\bm #1}}

\usepackage{multirow}
\usepackage{tabularx}
\newcolumntype{Y}{>{\centering\arraybackslash}X}
\usepackage{array}
\usepackage{makecell}

\begin{document}

\title{Hyperfine interaction in atomically thin transition metal dichalcogenides}

\author{I.~D.~Avdeev}
\author{D.~S.~Smirnov}
\email{smirnov@mail.ioffe.ru}
\affiliation{Ioffe Institute, 194021 St. Petersburg, Russia}

\date{\today}

\begin{abstract}
  Localization of charge carriers in monolayers (MLs) of transition metal dichalcogenides (TMDs) dramatically increases spin and valley coherence times, and, by analogy with other systems, the role of the hyperfine interaction should enhance. We perform theoretical analysis of the intervalley hyperfine interaction in TMD MLs based on the group representation theory. We demonstrate, that the spin-valley locking leads to the helical structure of the in-plane hyperfine interaction. In the upper valence band the hyperfine interaction is shown to be of the Ising type, which can be used for fabrication of the atomically thin quantum dots with the long spin and valley coherence times.
\end{abstract}

\maketitle

%====================================================
\section{Introduction}

Atomically thin TMDs, MX$_2$ with M being a transition metal (Mo, W) and X being a chalcogen (S, Se, T), represent a new generation of truly two-dimensional structures after graphene~\cite{novoselov04,novoselov05a}. 
Recently discovered~\cite{Mak2010,Splendiani2010}, they prove to have unique optical properties~\cite{MX2Review}: TMD MLs have a direct band gap at the two inequivalent $\bm K_+$ and $\bm K_-$ points of the Brillouin zone~\cite{Kormanyos2015,Yao_review}. 
The strong spin-orbit interaction leads to the pronounced spin splitting of the conduction and valence bands~\cite{Xiao2012,Kosmider2013,Molina2013} and results in the valley dependent optical selection rules: $\sigma^\pm$ light can induce optical transitions only in $K_\pm$ valleys, respectively~\cite{Cao2012,Xiao2012,Tarasenko_2D}. 
Generally, the optical properties of TMD MLs up to room temperature are determined by the Wannier-Mott excitons with the huge binding energy about $0.5$~eV~\cite{Cheiwchanchamnangij2012,Ramasubramaniam2012,Qiu2013,Chernikov2014,He2014,Wang2015}. 
These fascinating optical properties are believed to be potentially useful for a broad range of future applications~\cite{TMD_Electronics,Butler2013,Geim2013,Xia2014,Xu2014,Yu2015Valley,Castellanos2016,Mak2016}.

In recent years the interest shifted towards van der Waals heterostructures based on TMD MLs~\cite{Geim2013}.
However, novel opportunities are also opening in zero-dimensional systems, like quantum dots based on TMD MLs~\cite{Gui-Bin2014,MX2_hyperfine}. 
Such structures seem to be more appropriate for the applications in the optoelectronic devices~\cite{Chakraborty2015,He2015,Chakraborty2018,Atature2018} and potentially easier to realize. 
Indeed, any disorder in 2D structures leads to the localization of the charge carrier wavefunction~\cite{Lee1985,Kramer1993,Mirlin2008}, which increases the spin and valley coherence times.
In TMD MLs the charge carrier localization can be reached by means of the chemical exfoliation~\cite{Gopalakrishnan2015,Gan2015,Mukherjee2016}, lithographic nanopatterning~\cite{Wei2017}, wrinkles~\cite{Branny2016}, homojunctions~\cite{Li2018}, or defects~\cite{Koperski2015,Srivastava2015,Ganchev2015}.

In TMD MLs the spin and valley degrees of freedom are locked due to the strong spin-orbit coupling and exchange interaction in Coulomb complexes.
For excitons, the spin-valley polarization lifetime is limited by the exciton lifetime in the picosecond range. 
For resident charge carriers this limitation is released. 
The polarization can be preserved for a few nanoseconds in MoS$_2$~\cite{Yang2015,Yang2015nat,McCormick2018}, and even longer in WSe$_2$~\cite{Hsu2015,Song2016,Dey2017}. Particularly long polarization relaxation times can be obtained for the localized charge carriers, where the dominant role in the spin and valley dynamics is played by the hyperfine interaction with the host lattice nuclear spins~\cite{MX2_hyperfine,book_Glazov}.

The microscopic mechanisms of the spin relaxation were studied in very detail in the widespread A$_3$B$_5$ semiconductor quantum dots. 
It was found, that the main source of the spin relaxation in moderate magnetic fields is the hyperfine interaction with the randomly oriented host lattice nuclear spins~\cite{book_Glazov,Coisch_review}. We note, that the spin-orbit and the hyperfine interactions have the common relativistic origin, so the pronounced spin-orbit splitting of the conduction and valence bands in TMDs suggests the pronounced hyperfine interaction. It can be particularly strong in the case of the localization in a small area in the vicinity of a point defect.
Hence, we assume that the nuclear spin fluctuations are the dominant mechanism of the
polarization relaxation for the localized charge carriers in TMD MLs despite the moderate abundance of the nuclear isotopes with nonzero spin.

\begin{figure}
  \centering
  \includegraphics[width=0.8\linewidth]{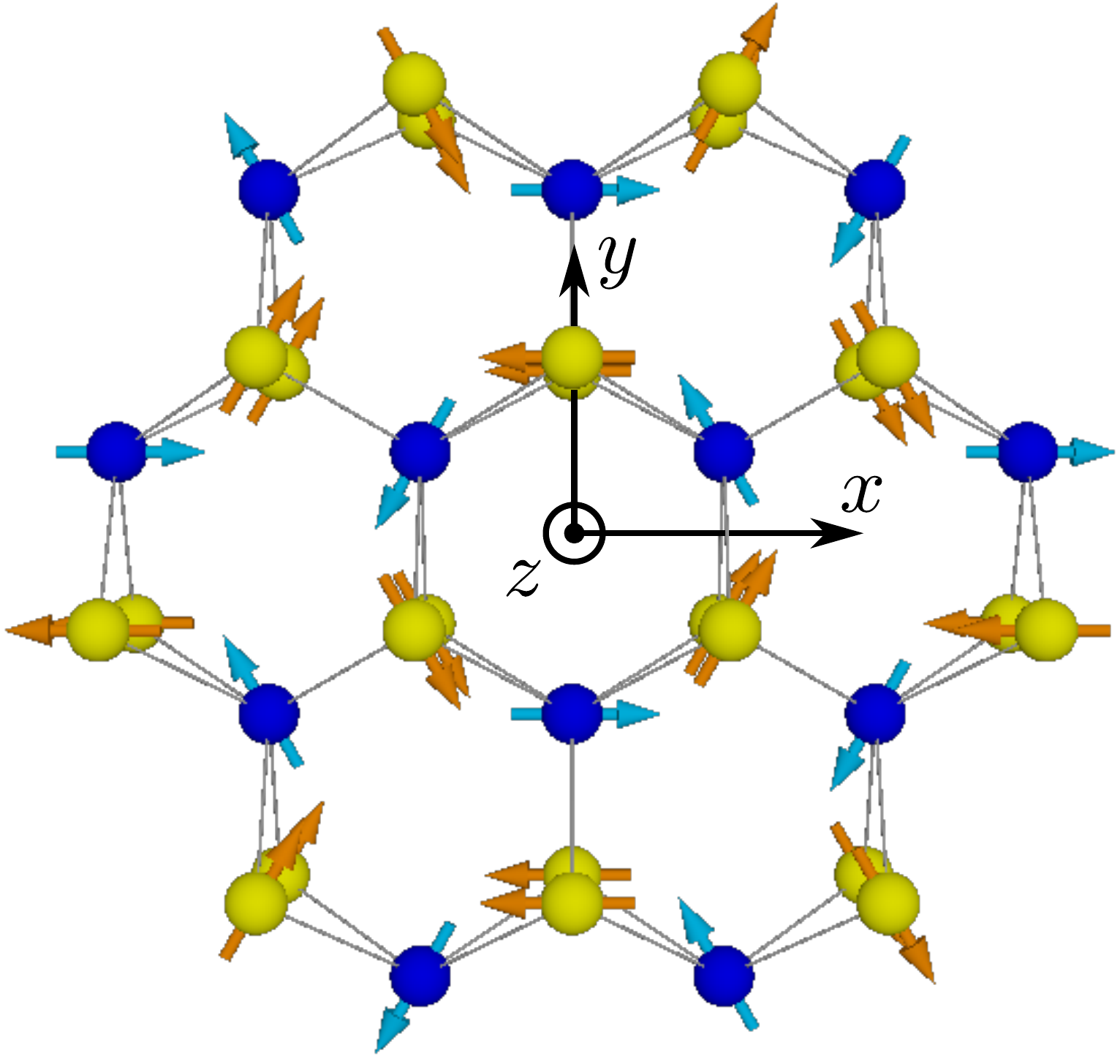}
  \caption{A part of TMD ML and the coordinate frame. The blue and yellow balls show the metal and chalcogen atoms, respectively, with arrows corresponding to the nuclear spins. The orientation of nuclear spins corresponds to the dynamic nuclear polarization induced by the valley pseudospin polarization along $x$ direction.}
  \label{fig:TMD}
\end{figure}

The spin-valley locking and specific band structure of TMDs can bring new challenges and peculiarities in the hardly explored field of the electron-nuclear interaction in TMD MLs~\cite{Sharma2017}. 
Any theoretical description of the effects related to the hyperfine interaction is based on the form of the hyperfine interaction Hamiltonian. In this work we derive it from the rigorous symmetry analysis. 
We also support our results with the microscopic analysis in the tight binding model.

In TMD MLs, the energy degenerate states in the $K_+$ and $K_-$ valleys are protected by the time reversal symmetry and the vertical mirror reflection symmetry of the structure. 
The nuclear spin polarization and fluctuations, similarly to an external magnetic field, can break both these symmetries. 
This results in the splitting and mixing of the degenerate states in the two valleys. 
Microscopically, the short-range nature of the hyperfine interaction favours the scattering of the charge carriers between the valleys~\cite{Tarasenko_hyperfine}. 
These effects are also qualitatively discussed in our work on the basis of the form of the hyperfine interaction Hamiltonian.

The paper is organized as follows. 
In Sec.~\ref{sec:symm} we perform the symmetry analysis of the hyperfine interaction.
Then, in Sec.~\ref{sec:micro}, using the tight binding model we calculate the components of the hyperfine interaction tensors. We provide estimates and discuss the results in Sec.~\ref{sec:disc}. Finally, conclusions are given in Sec.~\ref{sec:concl}.

%====================================================
\section{Symmetry analysis}
\label{sec:symm}

A single TMD ML has the honeycomb lattice with one transition metal atom and two chalcogen atoms in the unit cell, see Appendix~\ref{app:honeycomb}. 
A few unit cells of TMD ML are shown in Fig.~\ref{fig:TMD}, where the metal and chalcogen atoms are represented by blue and yellow balls, respectively. 
The metal atoms lie within the ML plane, $(xy)$, while the chalcogens are shifted along the $z$ axis in the opposite directions from the ML plane.

We chose the origin of the coordinate frame at the center of the hexagon, formed by the metal and chalcogen atoms~\cite{Kormanyos2015,MX2Review}, as shown in Fig.~\ref{fig:TMD}. 
We also choose the $y$ axis to be oriented towards the nearest pair of the chalcogen atoms. 
We note, that the caution should be taken to the choice of the coordinate frame origin and axes orientation, when comparing with the results of different authors~\cite{Fang2015,Yao_review}.

The point symmetry of the TMD ML is $D_{3h}$. 
This group consists of the horizontal (lateral) reflection plane $\sigma_h\parallel(xy)$, three fold rotation axis $C_3\parallel z$, three vertical reflection planes $3\sigma_v$, three in-plane two fold rotation axes $3C_2'$ (including $y$ axis), and the combinations $S_3=\sigma_h C_3$. In total, there are 12 symmetry operations including identity.

The valence and conduction band extrema are located in the two inequivalent $\bm K_{\pm}$ points of the Brillouin zone, see Appendix~\ref{app:honeycomb} for details. 
The wave vector point symmetry in these valleys is $C_{3h}$, which is a subgroup of $D_{3h}$ lacking all the elements interchanging the $\bm K_+$ and $\bm K_-$ valleys.

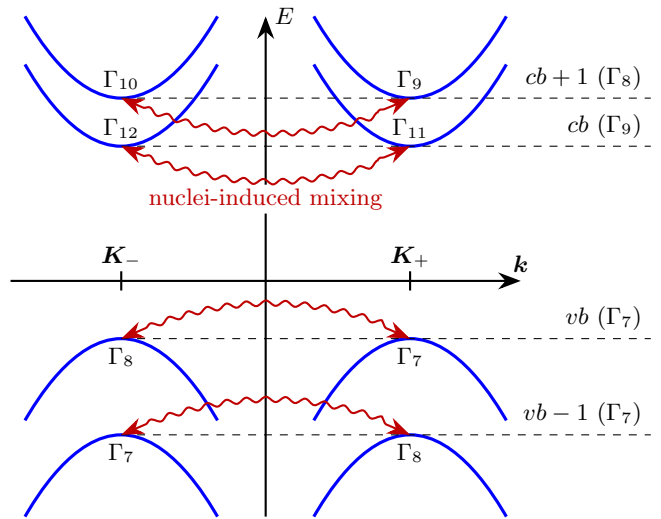
\begin{figure}
  	\centering{
	\begin{tikzpicture}[scale=0.64, >={Stealth[scale=1.3]}]
	
	% valley and k axis
	\newcommand\kaxy{-.8} % k axis y
	\newcommand\kmx{-3}
	\newcommand\kpx{+3}

	% right labels
	\newcommand\rlabx{3}
	
	% parabola parameters
	\newcommand\parwidth{4}
	\newcommand\parheightC{1.7}
	\newcommand\parheightV{1.7}
	\newcommand\mixheight{.7} % snaky arrows
	
	% bands y
	\newcommand\cbpy{3}
	\newcommand\cby{2}
	\newcommand\vby{-2}
	\newcommand\vbmy{-4}

		% K- scope
		\begin{scope}[shift={(\kmx,0)}]
			% cb+1
			\draw[very thick, blue] (-\parwidth/2, \cbpy+\parheightC) parabola bend (0, 3) (\parwidth/2, \cbpy+\parheightC);
			\node[above] at (0, \cbpy) {$\Gamma_{10}$};
		
			% cb
			\draw[very thick, blue] (-\parwidth/2, \cby+\parheightC) parabola bend (0, 2) (\parwidth/2, \cby+\parheightC);
			\node[above] at (0, \cby) {$\Gamma_{12}$};
		
			% vb
			\draw[very thick, blue] (-\parwidth/2, \vby-\parheightV) parabola bend (0,-2) (\parwidth/2,\vby-\parheightV);
			\node[below] at (0,\vby) {$\Gamma_{8}$};
		
			% vb - 1
			\draw[very thick, blue] (-\parwidth/2, \vbmy-\parheightV) parabola bend (0,-4) (\parwidth/2,\vbmy-\parheightV);
			\node[below] at (0,\vbmy) {$\Gamma_{7}$};
		\end{scope}
		
		% K+ scope
		\begin{scope}[shift={(\kpx,0)}]
			% cb+1
			\draw[very thick, blue] (-\parwidth/2, \cbpy+\parheightC) parabola bend (0, 3) (\parwidth/2, \cbpy+\parheightC);
			\node[above] at (0, \cbpy) {$\Gamma_{9}$};
		
			% cb
			\draw[very thick, blue] (-\parwidth/2, \cby+\parheightC) parabola bend (0, 2) (\parwidth/2, \cby+\parheightC);
			\node[above] at (0, \cby) {$\Gamma_{11}$};
		
			% vb
			\draw[very thick, blue] (-\parwidth/2, \vby-\parheightV) parabola bend (0,-2) (\parwidth/2,\vby-\parheightV);
			\node[below] at (0, \vby) {$\Gamma_{7}$};
		
			% vb - 1
			\draw[very thick, blue] (-\parwidth/2, \vbmy-\parheightV) parabola bend (0,-4) (\parwidth/2,\vbmy-\parheightV);
			\node[below] at (0, \vbmy) {$\Gamma_{8}$};
		\end{scope}

%		\draw[dashed, red] (-3, 1) -- (3, 1);
		\draw[thick, ->] (\kmx - \parwidth/2 - .3, \kaxy) -- (\kpx + \parwidth/2 + .3, \kaxy) node[above]{$\vec{k}$};
		\draw[thick] (0, \vbmy-\parheightV) -- (0, \cby-\mixheight-0.7);
		\draw[thick, ->] (0, \cby-\mixheight-0.1) -- (0, \cbpy+\parheightC) node[right]{$E$};		
		\draw[thick]
			(\kmx, \kaxy-.2) -- (\kmx, \kaxy+.2) node[above=-2]{$\vec{K}_-$}
			(\kpx, \kaxy-.2) -- (\kpx, \kaxy+.2) node[above=-2]{$\vec{K}_+$}
		;
		
		\draw[dashed] (\kmx, \cbpy) -- (\kpx+\parwidth/2+\rlabx, \cbpy) node[above left]{$cb+1$ ($\Gamma_8$)};
		\draw[dashed] (\kmx, \cby) -- (\kpx+\parwidth/2++\rlabx, \cby) node[above left]{$cb$ ($\Gamma_9$)};
		\draw[dashed] (\kmx, \vby) -- (\kpx+\parwidth/2++\rlabx, \vby) node[above left]{$vb$ ($\Gamma_7$)};
		\draw[dashed] (\kmx, \vbmy) -- (\kpx+\parwidth/2++\rlabx,\vbmy) node[above left]{$vb-1$ ($\Gamma_7$)} ;
		
		% some AUX information
		\begin{scope}[decoration={snake, amplitude=1,segment length=7.4}, line around/.style={decoration={pre length=#1,post length=0.8*#1}}]
			\draw[red!75!black, thick, decorate, line around=9, <->] (\kmx, \cbpy) parabola bend (\kmx/2+\kpx/2, \cbpy-\mixheight) (\kpx, \cbpy);
			\draw[red!75!black, thick, decorate, line around=9, <->] (\kmx, \cby) parabola bend (\kmx/2+\kpx/2, \cby-\mixheight) (\kpx, \cby);
			\node[below, red!75!black] at (0, \cby-\mixheight) {nuclei-induced mixing};
			% \node[below, red!75!black, fill=white] at (0, 1.5) {nuclei-induced mixing};
			% \node[below left, red!75!black] at (2.6, 1.5) {nuclei-induced mixing};
			\draw[red!75!black, thick, decorate, line around=10, <->] (\kmx,\vby) parabola bend (\kmx/2+\kpx/2,\vby+\mixheight) (\kpx,\vby);
			\draw[red!75!black, thick, decorate, line around=10, <->] (\kmx,\vbmy) parabola bend (\kmx/2+\kpx/2,\vbmy+\mixheight) (\kpx,\vbmy);
%			\draw[red!75!black, thick, decorate, line around=6, <->, decoration={markings,mark=at position 1 with
%    {\arrow[scale=3,>=stealth]{>}}},postaction=decorate] (-3,-3) parabola bend (0,-2.5) (3,-3);
		\end{scope}
		
	\end{tikzpicture}
	}
%%% Local Variables:
%%% mode: latex
%%% TeX-master: "Hyperfine_28Sep18_ds"
%%% End:
  \caption{Schematics of the band structure with the red wavy arrows showing the mixing between the energy degenerate states in $K_+$ and $K_-$ valleys induced by the hyperfine interaction.}
  \label{fig:bands}
\end{figure}

All the irreducible representations of the $C_{3h}$ group are one dimensional, so that all the electronic states in $\bm K_\pm$ valleys are nondegenerate, see the band diagram in Fig.~\ref{fig:bands}.
However, the two valleys are related by the time reversal symmetry and their energies coincide, in agreement with the Kramers theorem~\cite{ll3_eng}.

We focus our attention on the four bands in the vicinity of the band gap, which we will distinguish by the index $m=cb,\,cb+1,\,vb,\,vb-1$, as shown in Fig.~\ref{fig:bands}.
The electronic wave function in $K_{\pm}$ valley in the $m$th band is a Bloch function
\begin{equation}
  \label{eq:Bloch}
  \Psi_{\pm}^{(m)}(\bm r)=\e^{\i\bm K_{\pm}\bm r} u_{\pm}^{(m)}(\bm r),
\end{equation}
where $u_{\pm}^{(m)}(\bm r)$ is the Bloch amplitude (a spinor). Note, that the functions $\Psi_{+}^{(m)}(\bm r)$ and $\Psi_{-}^{(m)}(\bm r)$ are related by the time reversal symmetry or $\sigma_v$ reflection.

The nuclear spins weakly break the translation symmetry of the structure and lead to the splitting and mixing of the states in $\bm K_+$ and $\bm K_-$ valleys. 
The hyperfine interaction constants are usually of the order of $1~\mu$eV~\cite{MX2_hyperfine}. 
This is much smaller, than the spin-orbit splittings of the conduction and valence bands in TMD MLs, which are of the order of a few tens of meV and a few hundreds of meV, respectively~\cite{Kormanyos2015,DurnevUFN}. 
Therefore, the hyperfine interaction can mix only the states with the same energy, i.e. in the same band, as shown by the wavy arrows in Fig.~\ref{fig:bands}.

The two states in $K_+$ and $K_-$ valleys can be interpreted as a valley qubit~\cite{TMD_Electronics,Tarasenko_2D,Novoselov2D,2DValleytronics}. We introduce the valley pseudospin matrices $\hat{\bm\tau}=(\hat{\tau}_x,\hat{\tau}_y,\hat{\tau}_z)$, see Appendix~\ref{app:tau_op}, so that the $K_\pm$ states correspond to $\tau_z=\pm1/2$, respectively. Taking into account the form of the electronic wave functions, Eq.~\eqref{eq:Bloch},  the hyperfine interaction Hamiltonian in the $m$th band is
\begin{equation}
  \label{eq:Ham}
  \hat{\mathcal H}^{(m)}_{\rm{hf}}=\sum_n\e^{-\i\hat{\bm{K}}\bm{R}_n}\hat{\bm\tau}\hat{\bm A}_n^{(m)}\bm I_n\e^{\i\hat{\bm{K}}\bm{R}_n},
\end{equation}
where $n$ enumerates the nuclei with the spins $\bm I_n$ and the two dimensional coordinates $\bm R_n$, $\hat{\bm K}=2\bm K_+\hat{\tau}_z$ is the momentum operator and $\hat{A}_n^{(m)}$ is the tensor of the hyperfine interaction constants. 
Here we explicitly wrote the two operators (or matrix exponents) $\e^{\pm\i\hat{\bm{K}}\bm{R}_n}$, which account for the relative spatial phase shift between $\Psi_{+}^{(m)}(\bm r)$ and $\Psi_{-}^{(m)}(\bm r)$. As a result, the tensors $\hat{A}_n^{(m)}$ are independent of the position of the elementary cell in the TMD ML. Moreover, the hyperfine interaction tensors for each pair of chalcogen atoms in the same unit cell are related by the reflection in the horizontal plane, i.e. they are linearly dependent. As a result there are only two independent hyperfine interaction tensors in each band: one with metal and one with chalcogen atoms.

Further symmetry analysis allows one to find the restrictions on the form of the hyperfine interaction tensors. 
It is most convenient to start the analysis from the $C_{3h}$ group of the wave vector, and then consider the raise of the symmetry up to the $D_{3h}$, the point symmetry of the structure.

Each single nuclear spin $\bm I_n$, with the lateral coordinates $\vec{R}_n=(R_{n,x},R_{n,y})$, breaks the translation symmetry of the structure, which allows for the mixing of the states in the two $K_{\pm}$ valleys.
To analyze the hyperfine interaction tensor with the $n$th nucleus, it is convenient to move the origin of the coordinate frame towards the corresponding nucleus (by the two-dimensional vector $\vec{R}_n$). 
Upon this nontrivial translation, the irreducible representations of the wavefunctions $\Psi_\pm^{(m)}$ change. 
In Appendix~\ref{app:shift} we show, that for each symmetry operation $g$ of the wave vector point group $C_{3h}$ the matrix of the representation should be multiplied by
  \begin{equation}
    \label{eq:exp_D}
    e^{-i\vec{K}_{\pm}(\vec{R}_n - g \vec{R}_n)}.
  \end{equation}
Sets of these factors form representations, corresponding to the functions
\begin{equation}
  f_{\pm}(\vec{r})=\e^{-\i{\bm K}_{\pm}{\bm r}}\sum_{\bm a}\delta\left(\bm R_n+\bm a -\bm r\right).
  \label{eq:exp}
\end{equation}
Here the sum runs over all translation vectors $\bm a$ and $\delta(\bm r)$ is the Dirac $\delta$-function. The functions $f_{\pm}$ transform according to the $\Gamma_{2,3}$ ($\Gamma_{3,2}$) representations of the $C_{3h}$ point symmetry group for the $n$th atom being a metal (chalcogen), respectively. Therefore,  representations of the wavefunctions for the two different origins of the coordinate frame are related simply by the multiplication by the representation $\Gamma_2$ or $\Gamma_3$.

\begin{table}
      \caption{Spinor irreducible representations of the electronic states in the $\bm K_{\pm}$ valleys for different choices of the coordinate frame origin: in the center of hexagon (O), at metal atom (M), and between neighboring chalcogen atoms (X) for the $C_{3h}$ point group. The order of bands corresponds to the Mo-based structures.}
      \label{tab:groups}
        \begin{tabularx}{\linewidth}{c |YY | YY | YY}
          \hline\hline
			\multirow{2}{*}{band} & \multicolumn{2}{c|}{O} & \multicolumn{2}{c|}{M} & \multicolumn{2}{c}{X} \\
			\cline{2-7}
			& $\vec{K}_+$ & $\vec{K}_-$ & $\vec{K}_+$ & $\vec{K}_-$ & $\vec{K}_+$ & $\vec{K}_-$ \\
			\hline
$cb+1$ & $\Gamma_{9}$ & $\Gamma_{10}$ & $\Gamma_{8}$ & $\Gamma_{7}$ & $\Gamma_{12}$ & $\Gamma_{11}$ \\
$cb$ & $\Gamma_{11}$ & $\Gamma_{12}$ & $\Gamma_{7}$ & $\Gamma_{8}$ & $\Gamma_{10}$ & $\Gamma_{9}$ \\
$vb$& $\Gamma_{7}$ & $\Gamma_{8}$ & $\Gamma_{10}$ & $\Gamma_{9}$ & $\Gamma_{11}$ & $\Gamma_{12}$ \\
$vb-1$& $\Gamma_{8}$ & $\Gamma_{7}$ & $\Gamma_{12}$ & $\Gamma_{11}$ & $\Gamma_{9}$ & $\Gamma_{10}$ \\
\hline\hline
        \end{tabularx}
\end{table}

In Table~\ref{tab:groups} we present the irreducible representations of electronic states in $C_{3h}$ group in the bands under study ($cb+1$, $cb$, $vb$ and $vb-1$). 
The representations corresponding to the standard choice of the origin of the coordinate frame at the center of the hexagon are well established~\cite{Wang2017,MX2Review}. 
The representations corresponding to the shifted origin of the coordinate frame can be calculated using the multiplication rules for the $C_{3h}$ group~\cite{koster63}, as described above.
The two representations, corresponding to the two valleys in the same band, are always conjugate, in agreement with the time reversal symmetry. 
They join in $D_{3h}$ group in pairs~\cite{Wang2017,koster63}
\begin{subequations}
  \label{eq:phases}
  \begin{equation}
	\left\lbrace \Gamma_8(C_{3h}),\,\Gamma_7(C_{3h})\right\rbrace\to\Gamma_7(D_{3h}),
  \end{equation}
  \begin{equation}
    \left\lbrace \Gamma_{10}(C_{3h}),\,\Gamma_9(C_{3h})\right\rbrace\to\Gamma_8(D_{3h}),
  \end{equation}
  \begin{equation}
    \left\lbrace \Gamma_{11}(C_{3h}),\,\Gamma_{12}(C_{3h})\right\rbrace\to\Gamma_9(D_{3h}).
  \end{equation}
\end{subequations}
The order of the representations in the curly brackets corresponds to the standard basis of the representation to the right of the arrow~\cite{koster63}. This rule, together with Table~\ref{tab:groups}, allows one to find the irreducible representations of the pairs of wavefunctions $\Psi_\pm^{(m)}$ in the $D_{3h}$ point symmetry group with the coordinate frame origin being at $\vec{R}_n$. The results are listed in the second columns of Tables~\ref{tab:M} and~\ref{tab:X}. One can see, that it can be either $\Gamma_7$, or $\Gamma_8$, or $\Gamma_9$. For the rest of the paper all irreducible representation labels will refer to the $D_{3h}$ point group.

The irreducible representations corresponding to the components of the valley pseudospin $\hat{\bm \tau}$ can be found from the decomposition of the squares of the self-conjugate representations found above. The multiplication rules read~\cite{koster63}
\begin{subequations}
  \label{eq:mult}
  \begin{equation}
    \label{eq:mult78}
    \Gamma_7\otimes\Gamma_7=\Gamma_8\otimes\Gamma_8=\Gamma_1\oplus\Gamma_2\oplus\Gamma_5,
  \end{equation}
  \begin{equation}
    \label{eq:mult9}
    \Gamma_9\otimes\Gamma_9=\Gamma_1\oplus\Gamma_2\oplus\Gamma_3\oplus\Gamma_4.
  \end{equation}
\end{subequations}
Specifically, in Appendix~\ref{app:tau_op} we show that for the representations $\Gamma_7$ and $\Gamma_8$ the valley pseudospin component $\hat{ \tau}_z$ transforms according to $\Gamma_2$, while $\hat{ \tau}_{x,y}$ form the basis of the representation $\Gamma_5$. For the representation $\Gamma_9$ we find, that $\hat{ \tau}_z$ again transforms according to $\Gamma_2$, while $\hat{ \tau}_x$ and $\hat{ \tau}_y$ form the bases of the representations $\Gamma_3$ and $\Gamma_4$, respectively.

Now let us turn to the classification of the nuclear spin components. 
We recall, that we perform the symmetry analysis in the point symmetry group $D_{3h}$ with the center of transformations chosen at $\bm R_n$ (a two dimensional atomic coordinate). 
The nuclear spin of a metal atom, $\vec{I}_M$, is a pseudovector, its components transform according to $\Gamma_2$ ($I_{M,z}$) and $\Gamma_5$ ($\pm I_{M,x} -\i I_{M,y}$) irreducible representations. The two chalcogen atoms at the two-dimensional coordinate $\bm R_n$ (in the same unit cell) exchange their places under the reflection in the horizontal plane $\sigma_h$. So it is useful to introduce the linear combinations of their spins $\bm I_X=\left(\bm I^{\rm up}+\bm I^{\rm down} \right)/2$  and  $\Delta\bm I=(\bm I^{\rm up}-\bm I^{\rm down})$, where the superscripts ``$\rm up$'' and ``$\rm down$'' refer to the spins of the atoms above and below $(xy)$ plane, respectively. 
Under the reflection $\sigma_h$ the components of $\bm I_X$ transform in the same way as those of $\bm I_M$, while the components of $\Delta\bm I$ additionally change the sign. Under the reflection in the vertical plane $\sigma_v$ the components of both $\bm I_X$ and $\Delta \bm I$ transform in the same way as those of $\bm I_M$. As a result the components $I_{X,z}$ and $\pm I_{X,x}-\i I_{X,y}$ belong to $\Gamma_2$ and $\Gamma_5$ representations, respectively; the component $\Delta I_z$ belongs to $\Gamma_3$ representation, while the others, $\Delta I_{x}$ and $\Delta I_{y}$, form the basis of the representation $\Gamma_6$.

\begin{table*}[tb]
  \begin{minipage}{.48\textwidth}
      \vspace{-0.3cm} 
      \caption{Irreducible representations of the wavefunctions in the point symmetry group $D_{3h}$ with the center of transformations at a metal atom and the components of the hyperfine interaction tensor with the metal atoms $\hat{\bm A}^M$.}
      \label{tab:M}
      \begin{ruledtabular}
        \begin{tabular}{cccccc}
          band & irrep(M) & $A_{xx}^M/A^M$ & $A_{yy}^M/A^M$ & $A_{zz}^M/A^M$ \\
          \hline
          $cb+1$ & $\Gamma_7$ & $2/7$ & $-2/7$ & $-4/7$ \\
          $cb$ & $\Gamma_7$ & $2/7$ & $2/7$ & $4/7$ \\
          $vb$ & $\Gamma_8$ & $0$ & $0$ & $24/7$ \\
          $vb-1$ & $\Gamma_9$ & $0$ & $0$ & $32/7$ \\
        \end{tabular}
      \end{ruledtabular}
  \end{minipage}
  \hfill
  \begin{minipage}{.48\textwidth}
    \vspace{-0.3cm}
    \caption{Irreducible representations of the wavefunctions in the point symmetry group $D_{3h}$ with the center of transformations between two neighboring chalcogen atoms and the components of the hyperfine interaction tensor with the chalcogen atoms $\hat{\bm A}^X$.}
    \label{tab:X}
    \begin{ruledtabular}
      \begin{tabular}{cccccc}
        band & irrep(X) & $A_{xx}^X/A^X$ & $A_{yy}^X/A^X$ & $A_{zz}^X/A^X$ \\
        \hline
        $cb+1$ & $\Gamma_9$ & $0$ & $0$ & $-8/5$ \\
        $cb$ & $\Gamma_8$ & $-6/5$ & $6/5$ & $-12/5$ \\
        $vb$ & $\Gamma_9$ & $0$ & $0$ & $8/5$ \\
        $vb-1$ & $\Gamma_8$ & $-6/5$ & $-6/5$ & $12/5$ \\
      \end{tabular}
    \end{ruledtabular}
  \end{minipage}
\end{table*}

Now the symmetry analysis of the hyperfine interaction becomes straightforward. According to the method of invariants the coupling is allowed only between the components of $\hat{\bm \tau}$ and $\bm I$, which transform according to the same irreducible representation. For the representations $\Gamma_7$ and $\Gamma_8$ corresponding to the coordinate $\bm R_n$ of the $n$-th nucleus the hyperfine interaction Hamiltonian has the form:
  \begin{multline}
    \label{eq:H78}
    \mathcal H_{n}^{\Gamma_7,\Gamma_8}=A_{n,xx}\left(\hat{\tau}_x\cos\phi_n+\hat{\tau}_y\sin\phi_n\right)I_{n,x}\\+A_{n,yy}\left(\hat{\tau}_y\cos\phi_n-\hat{\tau}_x\sin\phi_n\right)I_{n,y}+A_{n,zz}\hat{\tau}_zI_{n,z},
  \end{multline}
where we took into account the phase factors in Eq.~\eqref{eq:Ham} $\phi_n=2KR_{n,x}$ with $K=|\bm K_{\pm}|$. The absolute values of in-plane hyperfine coupling constants are equal: $|A_{n,xx}|=|A_{n,yy}|$, while their signs can be opposite or the same depending on the representations of the electronic states in $K_+$ and $K_-$ valleys in the $C_{3h}$ group. Below we focus on the Mo-based structures, as for the W-based structures the results are the same, except for the swap of the band $cb$ with $cb+1$. Thus one has $A_{xx}=A_{yy}$ for Mo atoms in the $cb$ and for chalcogen atoms in $vb-1$. By contrast, for Mo atoms in the bands $cb+1$ and $vb$ and for chalcogen atoms in $cb$ one has $A_{yy}=-A_{xx}$, see Tables~\ref{tab:M} and~\ref{tab:X}.

In the case of the representation $\Gamma_9$ the Hamiltonian has the form
  \begin{multline}
    \label{eq:H9}
    \mathcal H_n^{\Gamma_9}=A_{n,zz}\hat{\tau}_zI_{n,z}\\+A_{n,xz}\left(\hat{\tau}_x\cos\phi_n+\hat{\tau}_y\sin\phi_n\right)\left(I_{n,z}^{up}-I_{n,z}^{down}\right).
  \end{multline}
Here in the second line $\bm I_{n}^{up,down}$ is the spin of chalcogen atom above (below) $(xy)$ plane, and this term is absent for Mo atoms. 
The Hamiltonian of this type is relevant for Mo atoms in the band $vb-1$ and for the chalcogen atoms in the bands $vb$ and $cb+1$.

One can see, that the component $A_{zz}$ is symmetry allowed for all atoms in all the bands. This means, that the nuclear spin polarization along $z$ lifts the Kramers degeneracy of the bands and splits the energy degenerate states, similarly to an external longitudinal magnetic field. The difference between $A_{zz}$ in the pairs of bands $cb,cb+1$ and $vb,vb+1$ is analogous to the longitudinal spin $g$ factor of the charge carriers~\cite{Mitioglu2015,Arora2016}. The in-plane hyperfine coupling is allowed only for the certain bands and atoms and has the ``helical'' structure, described by the phase factors in Eqs.~\eqref{eq:H78} and~\eqref{eq:H9}. In particular, from Eq.~\eqref{eq:H78} one can see, that the coupling coefficients between the valley pseudospin components and nuclear spin components periodically depend on the atomic coordinates. Moreover, the hyperfine interaction can be noncollinear, see the second line in Eq.~\eqref{eq:H9}. The coupling between the in-plane valley pseudospin components and out-of-plane nuclear spin components give rise to a number of interesting phenomena for localized charge carriers in TMD MLs~\cite{Urbaszek}.

The relations described above are strict and follow only from the symmetry analysis. The same results apply also for any substitutional impurity in TMD MLs.
Below we obtain additional restrictions on the components of the hyperfine interaction tensors from the microscopic analysis using the tight binding model. The results of the microscopic analysis are summarized in the Tables~\ref{tab:M} and~\ref{tab:X} and are in full agreement with the symmetry analysis performed in this section.

%====================================================
\section{Tight binding model}
\label{sec:micro}

The Hamiltonian of the hyperfine interaction with the nuclei has the form
\begin{equation}
  \hat{\mathcal H}_{\rm{hf}}=\sum_n2\mu_B\mu_I^{(n)}{\bm I}_n\left[\frac{8\pi}{3}\hat{\bm s}\delta(\bm r_n)+\frac{\hat{\bm l}_n}{r_n^3}-\frac{\hat{\bm s}}{r_n^3}+3\frac{{\bm r}_n\left(\hat{\bm s}{\bm r}_n\right)}{r_n^5}\right].
  \label{eq:Hhf}
\end{equation} 
Here $\mu_B$ is the Bohr magneton, $\mu_I^{(n)}=g_I^{(n)}\mu_N$ is the nuclear magnetic moment with $g_I^{(n)}$ and $\mu_N$ being the nuclear $g$-factors and the nuclear magneton respectively, $\bm r_n$ is the electron distance to the $n$th nucleus (being a three dimensional vector), $\hat{\bm l}_n=-i\left[\bm r_n\times\bm\nabla\right]$ is the electron angular momentum, and $\bm s$ is the electron spin operator. 
The first term in square brackets in Eq.~\eqref{eq:Hhf} describes the Fermi contact interaction, which vanishes for all atomic orbitals except for the $s$-type. The other terms describe magnetic dipole-dipole interaction.

The hyperfine interaction of charge carriers with the host lattice nuclei is a short range interaction~\cite{Grncharova-nuclei,MX2_hyperfine}, so it can be conveniently studied within the tight binding approximation. For the sake of simplicity, we limit ourselves with $d$-orbitals at the metal atoms (thus neglecting $s$-orbitals~\cite{Durnev_MX2}) and $p$-orbitals at the chalcogen atoms~\cite{Fang2015,Kormanyos2015}. In this basis the orbital part of the wavefunction is even with respect to the reflection $\sigma_h$ ($z\to-z$). As a result the Knight fields acting on the two chalcogen atoms in the same unit cell are the same, and the component $A_{xz}$ vanishes in this model. We note, that admixture of $d$-orbitals odd along $z$ axis at the chalcogen atoms~\cite{Wang2017} can provide nonzero coupling between $\tau_x$ and $\Delta I_z$ and make the hyperfine interaction anisotropic. The similar effect takes place for the holes in GaAs quantum dots with $C_{3v}$ symmetry~\cite{Vidal2016}.

Bearing in mind the irreducible representations of the point groups with the centers of transformations at a metal atom or between neighboring chalcogen atoms (Table~\ref{tab:groups}), one can find the explicit form of the Bloch amplitudes straight from the tight-binding model~\cite{Fang2015,Kormanyos2015}. For Mo-based structures they read
\begin{subequations}
  \label{eq:us}
  \begin{equation}
	\label{eq:usa}
    u_{\pm}^{cb+1}(\bm r)=\left[\pm\mathcal D_0(\bm r)\pm\mathcal P_{\mp 1}(\bm r)\right]\ket{\downarrow/\uparrow},
  \end{equation}
  \begin{equation}
	\label{eq:usb}
    u_{\pm}^{cb}(\bm r)=\left[\mp\mathcal D_0(\bm r)\mp\mathcal P_{\mp1}(\bm r)\right]\ket{\uparrow/\downarrow},
  \end{equation}
  \begin{equation}
	\label{eq:usc}
    u_{\pm}^{vb}(\bm r)=\left[\mathcal D_{\pm2}(\bm r)+\mathcal P_{\pm1}(\bm r)\right]\ket{\uparrow/\downarrow},
  \end{equation}
  \begin{equation}
	\label{eq:usd}
    u_{\pm}^{vb-1}(\bm r)=\left[\mathcal D_{\pm2}(\bm r)+\mathcal P_{\pm1}(\bm r)\right]\ket{\downarrow/\uparrow}.
  \end{equation}
\end{subequations}
Here the functions $\mathcal D_{m_z}(\bm r)$ and $\mathcal P_{m_z}(\bm r)$ ($m_z=0,\pm1,\pm2$) denote the Bloch amplitudes formed by the $d$ and $p$ atomic orbitals at metal and chalcogen atoms respectively. 
In the vicinity of the $n$th nucleus the orbital wavefunction has the form 
\begin{equation}
  \label{eq:R}
  \mathcal R(\bm r_n)=R_n(\bm r_n) Y_{lm_z}(\theta,\phi),
\end{equation}
where the spherical harmonics $Y_{lm_z}(\theta,\phi)$ with $l=1,2$ correspond to the functions $\mathcal P_{m_z}$ and $\mathcal D_{m_z}$, respectively. The radial parts of the wavefunctions in Eq.~\eqref{eq:R} are normalized as
\begin{equation}
 \int\limits_0^\infty R_{n}^2(r)r^2\d r=1. 
\end{equation}
Each pair of functions~\eqref{eq:usa}---\eqref{eq:usd} forms the basis of an irreducible representation of the group $D_{3h}$, and the two corresponding functions belong to the conjugated irreducible representations of the group $C_{3h}$, in agreement with Eq.~\eqref{eq:phases}.
We also mention, that in Ref.~\onlinecite{MX2_hyperfine}, where the hyperfine interaction was studied from the first principles, the expression for $u_{\pm}^{vb}$ is different, which results in some discrepancies with our results.

Taking into account the explicit form of the wave functions, Eq.~\eqref{eq:us}, additional symmetry restrictions can be obtained for the hyperfine interaction tensors. 
Indeed, the Hamiltonian~\eqref{eq:Hhf} can not change the total electron angular momentum $f_z=m_z+s_z$ by more than $1$~\footnote{The last term in Eq.~\eqref{eq:Hhf} being a part of the dipole-dipole interaction can change the electron orbital momentum $m_z$ by $\pm2$, but simultaneously the electron spin $s_z$ changes by $\mp1$, so the total angular momentum $f_z$ can not be changed by more than 1, in agreement with the spherical symmetry of the Hamiltonian~\eqref{eq:Hhf}.}.
The $\Gamma_9$ representation corresponds to $f_z=\pm3/2$, so these states can not be mixed by the hyperfine interaction. 
In this case the components $A_{xx}$ and $A_{yy}$ vanish, in agreement with the general symmetry arguments. 
Additionally, the wavefunctions at the metal atoms in the upper valence band, $vb$, have the total angular momenta $f_z\pm5/2$, hence $A_{xx}=A_{yy}=0$, see Table~\ref{tab:M}. As a result the hyperfine interaction in the upper valence band is purely of the Ising type.

The calculation of the matrix elements of the Hamiltonian~\eqref{eq:Hhf} with the wavefunctions described by Eqs.~\eqref{eq:us} and~\eqref{eq:R} yields the relation between the in-plane and out-of-plane components of the hyperfine interaction tensors and their signs.
These results are summarized in the last three columns in Tables~\ref{tab:M} and~\ref{tab:X}. 
One can see, that $|A_{zz}|=2|A_{xx}|$, whenever $A_{xx}$ is nonzero. 
The absolute values of the hyperfine interaction constants in Tables~\ref{tab:M} and~\ref{tab:X} are determined by
\begin{equation}
  A^{M,X}=2\mu_B\mu_I^{M,X}\int_0^\infty \frac{R_{M,X}^2(r)}{r}\d r,
\end{equation}
where $M$ and $X$ stand for the metal and chalcogen atoms, respectively. Importantly, one can see, that $A^{M,X}>0$, so Tables~\ref{tab:M} and~\ref{tab:X} yield the overall signs of the hyperfine interaction constants. In particular, the sign of $A_{zz}$ can be both positive and negative in different bands.

%====================================================
\section{Discussion}
\label{sec:disc}

In TMDs not all the metal and chalcogen isotopes have nonzero nuclear spins. The ones with nonzero nuclear spins are listed in Table~\ref{tab:isotopes} together with their abundances ($\nu$) and spins ($I$). One can see, that less than a half of atoms of each type have nonzero spins.

The values of the hyperfine interaction constants ($A$) can be calculated using atomistic approaches, for example DFT~\cite{Overhof2004,Ramzi2015,MX2_hyperfine}. 
The orbitals $\mathcal D_{m_z}$ at molybdenum and tungsten are related mainly to the $4d$ and $5d$ atomic orbitals, respectively, while the orbitals $\mathcal P_{m_z}$ are related to the $3p$, $4p$ and $5p$ atomic orbitals at sulfur, selenium and tellurium, respectively. The estimations for the hyperfine coupling constants can be obtained from the corresponding values known for the free atoms~\cite{Koh1985235}, see Table~\ref{tab:isotopes}. 
As one could expect, separately for metal and chalcogen atoms, the heavier is the atom the stronger is the hyperfine interaction. 
The fact that the interaction with nuclei is stronger for the chalcogen atoms than for the metal atoms is related to the fact, that $p$ orbitals correspond to the smaller angular momentum and are more localized at the nuclei, than $d$ orbitals. As a result the hyperfine interaction is most pronounced for tellurium atoms. We note, that the spin-orbit splitting of the conduction and valence bands in TMD MLs qualitatively obeys the same rules. This is related to the common relativistic origin of the two effects.

\begin{table}[b]
      \caption{Properties of the isotopes of M and X atoms with nonzero spin.}
      \label{tab:isotopes}
      \begin{ruledtabular}
        \begin{tabular}{ccccc}
          & $M$ & $\nu$ (\%) & $I$ & $A$ ($\mu$eV) \\
          \hline
          \makecell{Mo \\ ~} & \makecell{95\\97} & \makecell{15.92 \\ 9.55} & \makecell{5/2 \\ 5/2} & 0.57 \\
          \makecell{W \\ ~} & \makecell{183\\186} & \makecell{14.31 \\ 28.43} & \makecell{1/2 \\ 3/2} & 0.64 \\
          S & 33 & 0.76 & 3/2 & 0.75 \\
          Se & 77 & 7.63 & 1/2 & 3.9 \\
          \makecell{Te \\ ~} & \makecell{123\\125} & \makecell{0.89 \\ 7.07} & \makecell{1/2 \\ 1/2} & 8.3
        \end{tabular}
      \end{ruledtabular}
\end{table}

It is instructive to compare the hyperfine interaction parameters with those in the well studied bulk semiconductor GaAs.
In GaAs the spin-orbit splitting of the valence band is about $330$~meV~\cite{Adachi}, which approximately equals to the splitting of the two uppermost valence bands in TMD MLs. The hyperfine interaction constants in the valence band of GaAs are of the order of $10~\mu$eV~\cite{Chekhovich_Hyperfine,eh_noise}, which is comparable to those in TMDs.
Contrary, the hyperfine interaction in the conduction band of GaAs is an order of magnitude stronger due to $s$ type of the Bloch amplitudes~\cite{PhysRevB.15.5780,Chekhovich_constants}. Hence the hyperfine interaction of electrons in TMD MLs is much weaker, than in GaAs.

The most important consequences of the hyperfine interaction are the valley and spin relaxation of localized charge carriers and the dynamic nuclear spin polarization. In accordance with the classical model of Merkulov, Efros and Rosen~\cite{merkulov02} the spin relaxation in zero magnetic field consists of two stages. In the first stage, the charge carrier spin precesses in the static fluctuation of the Overhauser field $\bm B_N$, while the nuclear spin dynamics can be neglected. During this stage the initial spin polarization along $z$ axis decreases on average by a factor $f$. 
In the second stage, the nuclear spin dynamics comes into play, and the charge carrier spin relaxes to zero at parametrically longer timescale.

For the isotropic hyperfine interaction (e.g. when the Fermi contact interaction dominates), the factor $f$ equals to $1/3$~\cite{merkulov02}. 
In case of TMD MLs the hyperfine interaction is anisotropic. 
The hyperfine interaction anisotropy can be described by the parameter $\lambda$, which is determined by the relations
\begin{equation}
  \label{eq:lambda}
  \lambda^2=\frac{\braket{B_{N,z}^2}}{\braket{B_{N,x}^2}}=\frac{\braket{B_{N,z}^2}}{\braket{B_{N,y}^2}},
\end{equation}
where the angular brackets denote the time or ensemble averaging. Hereafter we assume that the nuclear spins are (i) independent of each other, (ii) randomly oriented, and (iii) the number of nuclear spins in the localization area is large. Provided the valley pseudospin is initially oriented along $z$, it decreases due to the precession in $\bm B_N$ by a factor~\cite{book_Glazov}
\begin{multline}
  f(\lambda)=\left\langle{B_{N,z}^2}/{B_N^2}\right\rangle\\=\left\{\begin{array}{lr}
        \dfrac{\lambda^2\left(\sqrt{\lambda^2-1}-\arctan\sqrt{\lambda^2-1}\right)}{\left(\lambda^2-1\right)^{3/2}}, & \lambda>1\\
        1/3, & \lambda=1\\
        \dfrac{\lambda^2\left(-\sqrt{1-\lambda^2}+\arctanh\sqrt{1-\lambda^2}\right)}{\left(1-\lambda^2\right)^{3/2}}, & \lambda<1
        \end{array}\right.
      \:. 
\end{multline}
Particularly, one can see, that $f(1)=1/3$ and $f(0)=0$, as expected. 

As one can see from Tables~\ref{tab:M} and~\ref{tab:X}, in the lower conduction band $\lambda=2$, and
\begin{equation}
  f(2)=\frac{4}{27}\left(9-\sqrt{3}\pi\right)\approx\frac{1}{2}.
\end{equation}
Therefore, the valley polarization of localized charge carriers in this band decreases approximately two times in a few tens of nanoseconds before decaying to zero at longer times. A small admixture of $s$-type orbitals at metal atoms~\cite{Yao_review} can slightly change this ratio.

The most interesting situation in TMD MLs takes place in the upper valence band. Here, as follows from Tables~\ref{tab:M} and~\ref{tab:X}, $\lambda=\infty$ and $f=1$. This describes the situation, when the relaxation of the spin $z$ component is absent because of the Ising type of the hyperfine interaction. For chalcogen atoms this is the symmetry requirement, while for metal atoms this results from $\mathcal D_{\pm2}$ atomic orbitals, which can not be mixed by the hyperfine interaction Hamiltonian~\eqref{eq:Hhf}. Therefore, one should expect the longest valley coherence times for localized holes in TMD MLs.

At long timescales the valley relaxation of localized holes can be related with (i) two phonon processes~\cite{PhysRevB.64.125316,PhysRevB.66.161318,Kioseoglou2012}, (ii) single phonon processes in combination with the hyperfine interaction~\cite{Hyperfine-phonon}, or (iii) mixing of the energy degenerate states by the localization potential~\cite{Liu2013,Pearce2017,Szechenyi2018}. 
In the latter case, the localization potential should be atomically sharp, e.g. an impurity. 
Otherwise, the degree of mixing of the two valleys is proportional to the ratio $\alpha$ of the lattice constant and the localization length. 
This situation is similar to the one described above with effective degree of the hyperfine interaction anisotropy $\lambda_{\rm eff}\sim1/\alpha$.

Similarly, the dynamic nuclear spin polarization is inefficient in the upper valence band of TMD MLs, while in all the other bands it can be significant. 
Interestingly, due to the difference of the valleys wavevectors, $\vec{K}_+$ and $\vec{K}_-$, the Hamiltonian of the hyperfine interaction, Eqs.~\eqref{eq:H78} and~\eqref{eq:H9}, contains additional phase factors, that ``rotate'' nuclear spins. 
An example of nuclear spin polarization distribution for $\bm\tau\parallel\bm x$ is given in Fig.~\ref{fig:TMD}, where one can clearly see the ``helical'' structure of nuclear spin polarization. 
This situation can be realized under linearly polarized excitation of TMD MLs. 
Importantly, the in-plane dynamical nuclear spin polarization is nonuniform in space. 
Despite the nuclear spin ordering, on average, the nuclear spin polarization is absent, which is analogous to antiferromagnetic spin state~\cite{Baltz2018}. 
Accordingly, the uniform nuclear spin polarization in $(xy)$ plane, does not lead to the splitting or mixing of the valleys because of the difference of Bloch wavevectors $\bm K_+$ and $\bm K_-$.

Finally, we note, that in the case of the charge carrier localization in a two dimensional structure the many-body nuclear spin effects, such as nuclear spin self-polarization~\cite{self-polarization,korenev1999dynamic} and formation of nuclear spin polaron~\cite{NuclearPolaron,Oulton_polaron} can be pronounced. These effects can be used to additionally increase the valley pseudospin relaxation time and to realize robust control of its orientation, similarly to magnetic skyrmions~\cite{Jiang2015,Jiang2017}.

%====================================================
\section{Conclusion}
\label{sec:concl}

Using the symmetry and microscopic analysis we have obtained explicit expressions for the hyperfine interaction tensors in TMD MLs, Tables~\ref{tab:M} and~\ref{tab:X}. Particularly, the valley pseudospin flips induced by the nuclei can take place only for the total electron angular momentum $f_z=\pm1/2$ at the corresponding atom. This restriction leads to the Ising type of the hyperfine interaction in the upper valence band and favours long valley pseudospin relaxation times of localized holes. Dynamic nuclear spin polarization in TMD MLs is shown to be of the ferromagnetic or antiferromagnetic (``helical'') type for out-of- or in-plane valley pseudospin polarization, respectively.

%====================================================
\begin{acknowledgments}
We thank Q. Tong, W. Yao, and M. M. Glazov for stimulating and useful discussions and acknowledge the Basis Foundation. This work was supported by the Government of the Russian Federation (contract \#~14.W03.31.0011 at the Ioffe Institute of RAS).
\end{acknowledgments}

%====================================================
%====================================================
\appendix

%====================================================
\section{Summary of TMD ML structure}
\label{app:honeycomb}

Here we present a brief summary of the TMD ML geometry and the choice of the real and reciprocal lattices bases.

A single TMD ML has the honeycomb lattice with the two dimensional translation vectors $\vec{a}_1 = a_0 (1,0)$ and $\vec{a}_2 = a_0 (1/2, \sqrt3/2)$, where $a_0$ is the lattice constant. The choice of the coordinate frame is described in Sec.~\ref{sec:symm}. The metal atoms lie within the ML plane $(xy)$, while the chalcogen atoms are shifted from the plane along and opposite to $z$ axis. The position of the metal atom in the elementary cell is $\vec{\tau}= a_0/2(1, 1/\sqrt3 )$, and the two chalcogen atoms are located above and below the point $-\vec{\tau}$.

The reciprocal lattice is determined by the basis vectors $\vec{b}_{1}=2\pi/a_0(1, -1/\sqrt3)$ and $\vec{b}_2=2\pi/a_0(0, 2/\sqrt3)$. 
The extremes of the conduction and valence bands are located at the two inequivalent corners of the hexagonal Brillouin zone, i.e. in $\vec{K}_{\pm}= 2\pi/a_0(\pm2/3, 0)$ valleys.

\begin{table*}[tb]
  \begin{minipage}{.48\textwidth}
      \vspace{-0.3cm} 
      \caption{
From left to right: the bands; $C_{3h}$ irreducible representations of the electronic states in $\vec{K}_{\mp}$ valley with the transformations center at the midpoint of the hexagon and at a metal atom (M); the irreducible representations in $D_{3h}$ group and the basis functions (see Table~\ref{tab:c_chart}); the form of the hyperfine interaction Hamiltonian.}
      \label{tab:M_shift}
      \begin{tabularx}{\linewidth}{c || YY || YY || Y|YY |@{\hspace{0\arrayrulewidth}}c@{\hspace{0.7mm}}| c}
          \hline\hline
			\multirow{2}{*}{band} & \multicolumn{2}{c||}{$C_{3h}$} & \multicolumn{2}{c||}{$C_{3h}$, M} & \multicolumn{3}{c|}{$D_{3h}$, M} && \multirow{2}{*}{$H_{hf}$} \\
			\cline{2-8}
			& $\vec{K}_-$ & $\vec{K}_+$ & $\vec{K}_-$ & $\vec{K}_+$ & irrep & $\vec{K}_-$ & $\vec{K}_+$ && \\ 
			\hline
$cb+1$ & $\Gamma_{10}$ & $\Gamma_{ 9}$ & $\Gamma_{ 7}$ & $\Gamma_{ 8}$ & $\Gamma_7$ & $\psi_{2}^{7}$ & $\psi_{1}^{7}$ && $H_-$ \\
$cb$   & $\Gamma_{12}$ & $\Gamma_{11}$ & $\Gamma_{ 8}$ & $\Gamma_{ 7}$ & $\Gamma_7$ & $\psi_{1}^{7}$ & $\psi_{2}^{7}$ && $H_+$ \\
$vb$   & $\Gamma_{ 8}$ & $\Gamma_{ 7}$ & $\Gamma_{ 9}$ & $\Gamma_{10}$ & $\Gamma_8$ & $\psi_{2}^{8}$ & $\psi_{1}^{8}$ && $H_-$ \\
$vb-1$ & $\Gamma_{ 7}$ & $\Gamma_{ 8}$ & $\Gamma_{11}$ & $\Gamma_{12}$ & $\Gamma_9$ & $\psi_{1}^{9}$ & $\psi_{2}^{9}$ && $H_z$ \\
%       & $\Gamma_{ 9}$ & $\Gamma_{10}$ & $\Gamma_{12}$ & $\Gamma_{11}$ & $\Gamma_9$ & $\psi_{2}^{9}$ & $\psi_{1}^{9}$ && $H_z$ \\
%       & $\Gamma_{11}$ & $\Gamma_{12}$ & $\Gamma_{10}$ & $\Gamma_{ 9}$ & $\Gamma_8$ & $\psi_{1}^{8}$ & $\psi_{2}^{8}$ && $H_+$ \\
\hline\hline
        \end{tabularx}
  \end{minipage}
  \hfill
  \begin{minipage}{.48\textwidth}
    \vspace{-0.3cm}
      \caption{From left to right: the bands; $C_{3h}$ irreducible representations of the electronic states in $\vec{K}_{\mp}$ valley with the transformations center at the midpoint of the hexagon and between neighboring chalcogen atoms (X); the irreducible representations in $D_{3h}$ group and the basis functions (see table~\ref{tab:c_chart}); the form of the hyperfine interaction Hamiltonian.}
      \label{tab:X_shift}
        \begin{tabularx}{\linewidth}{c ||YY || YY || Y|YY |@{\hspace{0\arrayrulewidth}}c@{\hspace{0.7mm}}| c}
          \hline\hline
			\multirow{2}{*}{band} & \multicolumn{2}{c||}{$C_{3h}$} & \multicolumn{2}{c||}{$C_{3h}$, X} & \multicolumn{3}{c|}{$D_{3h}$, X} && \multirow{2}{*}{$H_{hf}$} \\
			\cline{2-8}
			& $\vec{K}_-$ & $\vec{K}_+$ & $\vec{K}_-$ & $\vec{K}_+$ & irrep & $\vec{K}_-$ & $\vec{K}_+$ && \\ 
			\hline
$cb+1$ & $\Gamma_{10}$ & $\Gamma_{ 9}$ & $\Gamma_{11}$ & $\Gamma_{12}$ & $\Gamma_9$ & $\psi_{1}^{9}$ & $\psi_{2}^{9}$ && $H_z$ \\
$cb$ & $\Gamma_{12}$ & $\Gamma_{11}$ & $\Gamma_{ 9}$ & $\Gamma_{10}$ & $\Gamma_8$ & $\psi_{2}^{8}$ & $\psi_{1}^{8}$ && $H_-$ \\
$vb$ & $\Gamma_{ 8}$ & $\Gamma_{ 7}$ & $\Gamma_{12}$ & $\Gamma_{11}$ & $\Gamma_9$ & $\psi_{2}^{9}$ & $\psi_{1}^{9}$ && $H_z$ \\
$vb-1$ & $\Gamma_{ 7}$ & $\Gamma_{ 8}$ & $\Gamma_{10}$ & $\Gamma_{ 9}$ & $\Gamma_8$ & $\psi_{1}^{8}$ & $\psi_{2}^{8}$ && $H_+$ \\
%       & $\Gamma_{ 9}$ & $\Gamma_{10}$ & $\Gamma_{ 8}$ & $\Gamma_{ 7}$ & $\Gamma_7$ & $\psi_{1}^{7}$ & $\psi_{2}^{7}$ && $H_+$ \\
%       & $\Gamma_{11}$ & $\Gamma_{12}$ & $\Gamma_{ 7}$ & $\Gamma_{ 8}$ & $\Gamma_7$ & $\psi_{2}^{7}$ & $\psi_{1}^{7}$ && $H_-$ \\
  \hline\hline
        \end{tabularx}
  \end{minipage}
\end{table*}

%====================================================
\section{Shift of point symmetry group transformations center}
\label{app:shift}

For the symmetry analysis, described in Sec~\ref{sec:symm}, it is necessary to find the irreducible representations of the electronic states for the choices of the point symmetry origins at the $\pm\vec{\tau}$ points. 
Here we present a formalism, which solves the problem. 
It can also be easily generalized for the other problems in physics of multivalley semiconductors.

Let us consider a spatial group $G_{\vec{k}}$ of the wave vector $\bm k=\bm K_\pm$. 
Further, let us assume, that this group is symmorphic, as in the case of TMD MLs. 
Let $T$ be the translation subgroup. 
Then the quotient group $P_{\vec{k}}=G_{\vec{k}}/T$, being the set of rotations $\{R\}$, is the corresponding point group. 
By definition, for each $R$
~\cite{birpikus_eng}
\begin{equation}
	R \, \vec{k} = \vec{k} + \vec{b} \:,
        \label{eq:Rk}
\end{equation}
where $\vec{b}$ is a reciprocal lattice vector (or zero vector).

Now consider a set of the Bloch wave functions $\{ \psi_n \} = \{ e^{i \vec{k} \vec{r}} u_n(\vec{r}) \}$, with $u_n(\vec{r})$ being the Bloch amplitude, that forms the basis of an irreducible representation $D$ of the point group $P_{\vec{k}}$. 
Note, that in case of TMD ML, all the irreducible representations are one dimensional, but here we consider a general situation. 
The wave functions transform under the action of $R$ as
\begin{equation}
  R\, \psi_n = \sum_{n'} \psi_{n'} \, D_{n'n}(R)
\:.
\label{eq:D}
\end{equation}
where $\hat{D}(R)$ are the matrices of the irreducible representation $D$.

We introduce the point group $P_{\vec{k}}'$ with the transformations center shifted by a vector $\vec{t}$ as
\begin{equation}
	P'_{\vec{k}} = \hat{\vec{t}} \, P_{\vec{k}} \, \hat{\vec{t}}^{-1},
\end{equation}
where $\hat{\vec{t}}$ denotes the translation operator. Note, that generally $\vec{t}$ is not a translation vector.
The elements $g'$ of $P_{\vec{k}}'$ have the form
\begin{equation}
  g' = (R | \vec{t} - R\,\vec{t}),
\end{equation}
being the rotation $R$ with the subsequent shift by the vector $\vec{t} - R\,\vec{t}$.

For TMD ML for $\vec{t}=\pm\vec{\tau}$, the vector $\vec{t} - R\,\vec{t}$ is a translation vector for each $R$ (see Appendix~\ref{app:honeycomb}), because there are only one metal atom and one pair of chalcogen atoms in the primitive cell. Once this condition is satisfied, one can show that the set $\{e^{-i\vec{k}(\vec{t}-R\vec{t})}\}$ forms a one dimensional representation $D_{\vec{t}}$ of the group $P_{\vec{k}}$.

To prove this statement let us consider the two rotations $R_1, R_2 \in P_{\vec{k}}$ and the corresponding translation vectors
  \begin{equation}
    \label{eq:alpha}
    \vec{\alpha}_{1,2}=\vec{t}-R_{1,2}\vec{t}\:.
  \end{equation}
  Since the scalar product is invariant under rotations $\vec{k}R_1\vec{\alpha}_2=\left(R_1^{-1}\vec{k}\right)\vec{\alpha}_2$, and
\[
\e^{ i \vec{k}\left(\vec{\alpha}_2-R_1\vec{\alpha}_2\right)}=1
\]
by the definition of $P_{\vec k}$, see Eq.~\eqref{eq:Rk}. From this relation, using Eq.~\eqref{eq:alpha}, we readily obtain
\[
e^{-i\vec{k} (\vec{t} - R_1 \vec{t}) } e^{-i\vec{k} (\vec{t} - R_2 \vec{t}) }=\e^{-\i\vec{k}\left(\vec{t}-R_1R_2\vec{t}\right)},
\]
which proves the multiplication rule, and thus the statement.

Therefore, the set of the functions $\lbrace\psi_n\rbrace$, which transforms according to Eq.~\eqref{eq:D}, forms the basis of a representation
\begin{equation}
	D' = D \otimes D_{\vec{t}} \:,
\end{equation}
of the point group $P_{\vec{k}}'$, where $D_{\vec{t}}(R) = e^{-i\vec{k}(\vec{t} - R \vec{t})}$, in agreement with Eq.~\eqref{eq:exp_D}.

In the $\vec{K}_-$ ($\vec{K}_+$) valley of a TMD ML, $D_{\vec{\tau}}$ is the $\Gamma_{2(3)}$ representation and $D_{-\vec{\tau}}$ is the representation $\Gamma_{3(2)}$ of the $C_{3h}$ point group. The multiplication of the representations allows us to find the irreducible representations of the wave functions with different centers of transformations. The representations of $D_{3h}$ group are summarized in Tables~\ref{tab:M} and~\ref{tab:X}, while the representation of $C_{3h}$ group are given in Tables~\ref{tab:M_shift} and~\ref{tab:X_shift}. The correspondence between them is established using the compatibility Table~\ref{tab:c_chart}.

\begin{table}[b]
%      \vspace{-0.3cm} 
      \caption{Compatibility chart of irreducible representations and basis functions for $D_{3h}$ and $C_{3h}$ point symmetry groups.}
      \label{tab:c_chart}
%      \begin{ruledtabular}
        \begin{tabularx}{\linewidth}{c|Y|Y||c|Y}
%        \{2}{c|}
          \hline\hline
		  \multicolumn{3}{c||}{$D_{3h}$} & \multicolumn{2}{c}{$C_{3h}$} \\
          \hline
		  $\Gamma_7$ & $\begin{array}{c} \psi_1^7 \\ \psi_2^7 \end{array}$ & $\begin{array}{c} \ket{-1/2} \\ \ket{+1/2} \end{array}$ & 
          $\begin{array}{c} \Gamma_{8} \\ \Gamma_7 \end{array}$ & $\begin{array}{c} \psi_1^7 \\ \psi_2^7 \end{array}$ \\ 
          \hline
          $\Gamma_8$ & $\begin{array}{c} \psi_1^8 \\ \psi_2^8 \end{array}$ & $\begin{array}{c} \ket{-1/2}zS_z \\ \ket{+1/2}zS_z \end{array}$ & 
          $\begin{array}{c} \Gamma_{10} \\ \Gamma_9 \end{array}$ & $\begin{array}{c} \psi_1^8 \\ \psi_2^8 \end{array}$ \\ 
          \hline
          $\Gamma_9$ & $\begin{array}{c} \psi_1^9 \\ \psi_2^9 \end{array}$ & $\begin{array}{c} \ket{-3/2} \\ \ket{+3/2} \end{array}$ & 
          $\begin{array}{c} \Gamma_{11} \\ \Gamma_{12} \end{array}$ & $\begin{array}{c} \psi_1^9 \\ \psi_2^9 \end{array}$ \\           
          \hline\hline
        \end{tabularx}
 %     \end{ruledtabular}
\end{table}

% =================================================================
\section{Pseudospin operators}
\label{app:tau_op}
Here we explicitly construct the pseudospin operators in the dyadic form, which act in the valley pseudospin $2\times2$ space. The bases of the representations are chosen in agreement with Ref.~\onlinecite{koster63}.

It follows from Eq.~\eqref{eq:mult78}, that in the corresponding cases one can construct the valley pseudospin operators belonging to $\Gamma_2$ and $\Gamma_5$ representations. We denote the standard bases of $\Gamma_{7}$ $(\Gamma_8)$ representation as $\vec{u} = (\ket{1}, \ket{2})$ and $\vec{v} = (-\bra{2}, \bra{1})$. These two bases can be obtain one from another performing the time reversal with the Hermitian conjugation. Thus they transform in the same way for all symmetry operations $g$:
  \begin{subequations}
    \begin{equation}
      g \, \vec{u} = \vec{u} \, \hat D(g),
    \end{equation}
    \begin{equation}
      g \, \vec{v} = \vec{v} \, \hat D(g).
    \end{equation}
  \end{subequations}
Using the coupling coefficients one can easily find operators belonging to $\Gamma_2$ representation
\begin{equation}
	\label{eq:tau_z_op}
	\hat{\tau}^{(2)} = \frac{i}{\sqrt2} (\ket{1}\bra{1} - \ket{2} \bra{2} ) \propto \hat{\tau}_z
	\:,
\end{equation}
and $\Gamma_5$ representation
\begin{equation}
	\begin{array}{l}
          \hat{\tau}^{(5)}_1 = -\ket{1}\bra{2} \propto \phantom{-} (\hat{\tau}_x - i \hat{\tau}_y) \:, \\ 
          \hat{\tau}^{(5)}_2 = \phantom{-} \ket{2} \bra{1} \propto -(\hat{\tau}_x + i \hat{\tau}_y) \:.
	\end{array}
\end{equation}
They are proportional to the components of the valley pseudospin operator $\hat{\vec{\tau}}$, because they transform in the same way. 
Finally, the intervalley hyperfine interaction Hamiltonian can be written in a matrix form as
\begin{equation}
	H_{\pm} = A_{\perp} (I_x \tau_x \pm I_y \tau_y ) + A_{\parallel} \tau_z \:,
\end{equation}
where $A_\perp$ and $A_\parallel$ are the constants and the different signs correspond to the different order of the basis functions $(\ket{1}, \ket{2})$ or $(\ket{2}, \ket{1})$. This order is given in the last columns of Tables~\ref{tab:M_shift} and~\ref{tab:X_shift}.

For $\Gamma_9$ representation of the wavefunctions the pseudovector operator $\hat{\tau}_z$ is given by Eq.~\eqref{eq:tau_z_op}. The operator $\hat{\tau}_x$ belonging to $\Gamma_3$ representation has the form
  \begin{equation}
    \hat{\tau}^{(3)}=\frac{1}{2}\left(\ket{1} \bra{2} + \ket{2} \bra{1}\right) = \hat{\tau}_x \:. \\
  \end{equation}
Note, that there is a misprint in Ref.~\onlinecite{koster63} in Table~67 with the corresponding coupling coefficients. 
Neglecting the noncollinear coupling between $\Delta I_z$ and $\tau_x$, the symmetry allowed hyperfine interaction Hamiltonian is
\begin{equation}
	\hat{H}_z = A_{\perp} I_z \hat{\tau}_z \:.
\end{equation}
The bands and atoms relevant for this case are also given in Tables~\ref{tab:M_shift} and~\ref{tab:X_shift} in the last column.

%\bibliography{all-1}


\end{document}